\documentclass[preprint, a4paper]{elsarticle}

\usepackage{graphicx}

\usepackage{amssymb}

\usepackage{float}
\restylefloat{table}

\journal{SoftwareX}

\usepackage{mathtools}
\usepackage[T1]{fontenc}
\usepackage{listings}
\usepackage{amsmath}
\usepackage{esint}
\usepackage[hyphens,spaces,obeyspaces]{url}
\usepackage{hyperref}
\hypersetup{
colorlinks=true,
linkcolor=blue,
filecolor=magenta,      
urlcolor=cyan,
}
\usepackage{cleveref} 
\crefname{figure}{fig.}{fig.} 
\Crefname{figure}{Fig.}{Fig.}
\crefname{equation}{eq.}{eq.} 
\Crefname{equation}{Eq.}{Eq.}
\usepackage{subcaption}
\usepackage{bm}
\usepackage{adjustbox}

\begin{document}

\begin{frontmatter}



\title{OpenPIV-Matlab - An open-source software for particle image velocimetry; test case: birds' aerodynamics}

\author[label1]{Hadar Ben-Gida}
\author[label2]{Roi Gurka}
\author[label3]{Alex Liberzon}

\address[label1]{Israeli Air Force, 6473428 Tel-Aviv, Israel}
\address[label2]{Department of Physics and Engineering Science, Coastal Carolina University, Conway, SC 29528, USA}
\address[label3]{School of Mechanical Engineering, Tel-Aviv University, Tel-Aviv, 6997801, Israel}

\begin{abstract}
    We present an open-source MATLAB package, entitled OpenPIV-Matlab, for analyzing particle image velocimetry (PIV) data. We extend the PIV analysis with additional tools for post-processing the PIV results including the estimation of aero/hydrodynamic forces from the PIV data of a wake behind an immersed (bluff or streamlined) body. The paper presents a detailed description of the packages, covering the three main parts: generating two-dimensional two component velocity fields from pairs of images (OpenPIV-Matlab), spatial and temporal flow analysis based on the velocity fields (Spatial and Temporal Analysis Toolbox), and wake flow analysis along with the force estimates (getWAKE Toolbox). A complete analysis with a variety of post-processing capabilities is demonstrated using time-resolved PIV wake data of a freely flying European starling (\emph{Sturnus vulgaris}) in a wind tunnel.
\end{abstract}


\end{frontmatter}




\begin{table}[H]
\begin{tabular}{|l|p{6.5cm}|p{6.5cm}|}
\hline
C1 & Current code version & v1.7 \\
\hline
C2 & Permanent link to code/repository used for this code version & \url{https://github.com/OpenPIV/openpiv-matlab},

\url{https://github.com/OpenPIV/openpiv-spatial-analysis-toolbox},

\url{https://github.com/OpenPIV/getWAKE}\\
\hline
C3 & Code Ocean compute capsule & \\
\hline
C4 & Legal Code License & MIT  \\
\hline
C5 & Code versioning system used & git \\
\hline
C6 & Software code languages, tools, and services used & MATLAB \\
\hline
C7 & Compilation requirements, operating environments \& dependencies & Linux, macOS, Windows, MATLAB (with Image Processing Toolbox, Statistics and Machine Learning Toolbox, Curve Fitting Toolbox and Signal Processing Toolbox) \\
\hline
C8 & If available Link to developer documentation/manual &  \url{https://github.com/OpenPIV/openpiv-matlab/blob/master/docs/Tutorial_OpenPIV/Tutorial_OpenPIV.pdf},

\url{https://github.com/OpenPIV/openpiv-spatial-analysis-toolbox/blob/master/docs/tutorial.rst},

\url{https://github.com/OpenPIV/getWAKE/blob/master/docs/\%E2\%80\%8F\%E2\%80\%8FgetWAKE-UsersManual.pdf}  \\
\hline
C9 & Support email for questions & \url{openpiv2008@gmail.com}\\
\hline
\end{tabular}
\caption{Code metadata}
\label{table1} 
\end{table}




\section{Motivation and significance}
\label{sec:motivation_significance}

Particle Image Velocimetry (PIV) is a state-of-the-art optical flow measurement technique. Using PIV, two- and three-dimensional (2D/3D) velocity fields are obtained with high spatial and temporal resolution, in a non-intrusive manner. PIV is widely used in research and industrial applications. For a detailed description of PIV principles and methodology, please refer to the book by Raffel et al.~\citep{raffel2018}. 

\begin{figure}[ht!]
	\centering
	\includegraphics[width=0.64\textwidth]{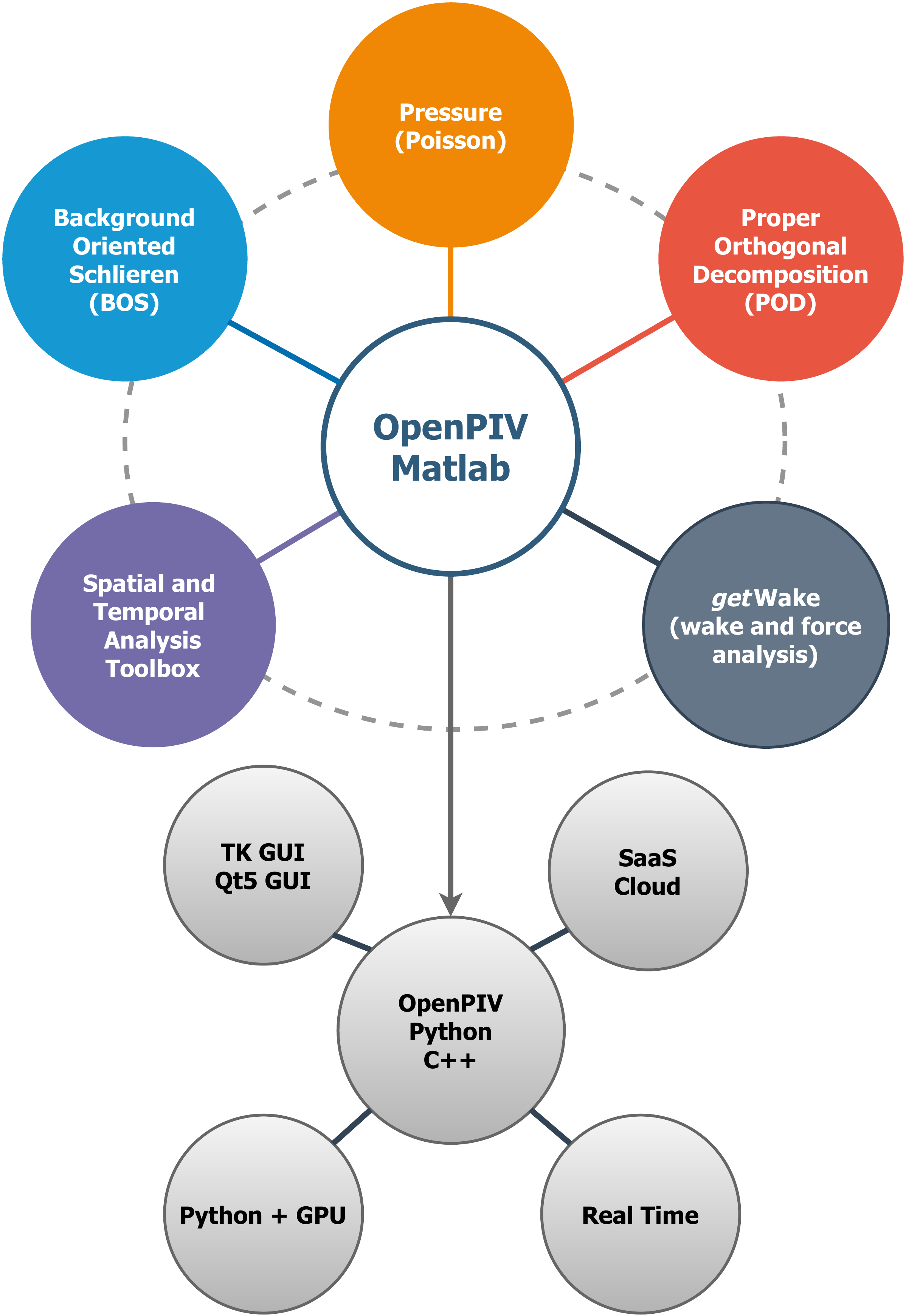}
	\caption{OpenPIV community software structure overview.}
	\label{fig:openpiv-full-map}
\end{figure}

OpenPIV is an international scientific community that develops free and open-source software that performs PIV analysis to obtain the velocity fields from images and a variety of post-processing tools to elucidate the physics of the investigated flow. OpenPIV houses several software packages, schematically shown in \Cref{fig:openpiv-full-map}. Originally, OpenPIV was developed to analyse PIV images to yield a 2D velocity field using Matlab platform~\cite{Gurka1999}. Later on, the spatial and temporal analysis toolbox was developed for the characterization of the measured flow field as well as provide turbulence characteristics and some features of spectral analysis based on PIV data. Two additional toolboxes were later added: Pressure and Proper Orthogonal Decomposition (POD)~\cite{Gurka1999,Gurka2006}. The pressure toolbox uses Poisson formulation of the flow equation to estimate the pressure field, assuming the boundary conditions of the field is known; using an iterative technique~\cite{Gurka1999}. The POD toolbox was developed to extract energetic modes from velocity or vorticity fields to study coherent motions in turbulent flows~\cite{Gurka2006,Liberzon2011j}. Another OpenPIV toolbox extended the post-processing capabilities with calculation of the density fields using the background-oriented schlieren technique~\cite{Verso2015}. Most recently, we added a toolbox to estimate aero/hydrodynamic loads based on PIV data acquired in a body wake, entitled getWAKE~\cite{BenGida2020}. 

In parallel to Matlab toolboxes, OpenPIV team developed also C++ and Python branches (see \Cref{fig:openpiv-full-map}). Detailed description of the C++ version can be found in Taylor et al.~\cite{Taylor2010}. The Python branch has additional features, including adaptive meshing, correlation of 3D-PIV images, real-time processing, among others. A paper of the OpenPIV-Python version is out of scope of this work and will be released separately.


The set of tools described hereinafter, provides an open-source "full-stack" package for analyzing and post-processing PIV data measured in the body wake. The significance is in the variety of tools, together providing a complete solution supporting new PIV users and experts that study aero/hydrodynamics based on wakes of dragged or self-propelled bodies.  




We demonstrate OpenPIV-Matlab capabilities using the PIV near wake flow data measured behind a freely flying European starling (\emph{Sturnus vulgaris}) in a wind tunnel~\cite{Kirchhefer2013} and present how we can shed light on the role of unsteady aerodynamics in natural flyers locomotion. The wake signature behind birds for estimating their aerodynamic performance is a very active field of research due to rapid technological advancement of high-speed lasers, cameras and data transfer solutions~\cite{Spedding2003,Johansson2009,BenGida2013,Gurka2017}.

The OpenPIV-Matlab stands-out due to the advanced analysis toolboxes with its capabilities beyond the other open-source PIV software, such as PIVlab~\cite{Thielicke2014}, MATPIV, among others.

\section{Software description}
\label{sec:software_description}

\subsection{Software Architecture}
\label{sec:software_architecture}

OpenPIV-Matlab uses the MATLAB (Mathworks Inc.) language to provide fast advanced PIV processing and post-processing tools and straightforward development process. 

We present here the three tools: OpenPIV (invoked using \texttt{opevpivgui.m}), Spatial and Temporal Analysis Toolbox (\texttt{spatialbox.m}) and getWAKE Toolbox (\texttt{wake.m}).

\begin{itemize}

\item \texttt{opevpivgui.m} is a GUI comprised of several subroutines that allow to import PIV images, pre-process them, analyse using fast Fourier transform-based cross-correlation algorithm, filter and interpolate the flow field, and export the velocity vector maps as ASCII files.

\item \texttt{spatialbox.m} is a GUI that allows the user to load a series of velocity maps created by \texttt{openpivgui.m} and calculate various flow characteristics (mean and turbulent, velocity derivatives, energy terms, auto-correlation, etc.), plot them as contours, vector fields and spatial distribution profiles and storing the processed data in Matlab MAT format. 

\item \texttt{wake.m} is a GUI which loads the MAT file exported by \texttt{spatialbox.m}. For the case of time-resolved wake data, it enables reconstruction of a complete wake signature behind the body using a cross-correlation algorithm that overlaps consecutive velocity maps. Based on the reconstructed wake data, \texttt{wake.m} also estimates aerodynamic body forces such as profile drag and cumulative circulatory lift. 

\end{itemize}

In the following we describe the functionality of each toolbox. 

\subsection{Software Functionalities}
\label{sec:software_functionalities}

\begin{figure}[ht!]
	\centering
	\includegraphics[width=0.85\textwidth]{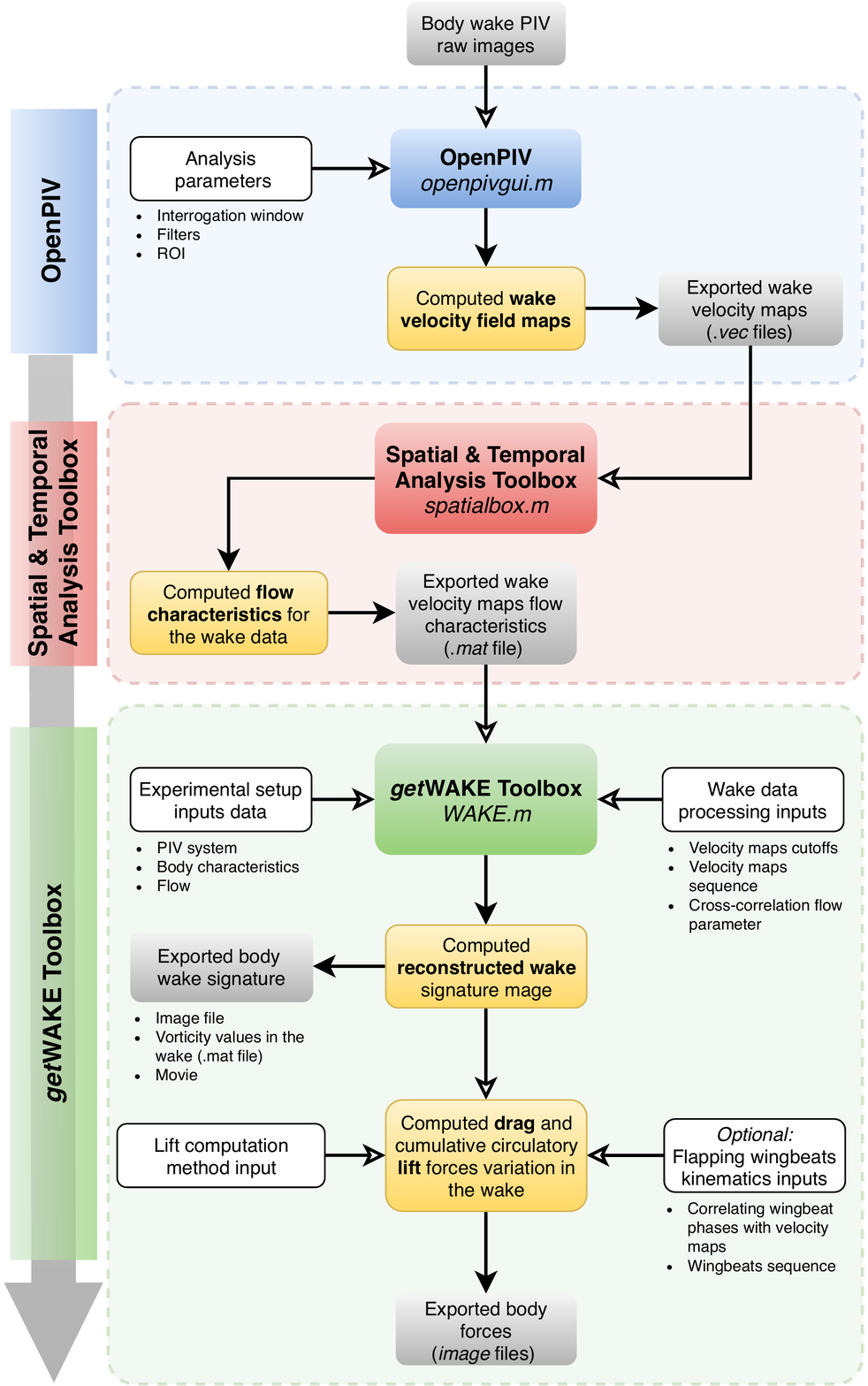}
	\caption{Schematic overview of the body PIV wake data analysis and post-processing routine in OpenPIV-Matlab.}
	\label{fig:openpiv-wake-routine}
\end{figure}

An example of a flowchart of analysis of the body wake PIV data is depicted in \Cref{fig:openpiv-wake-routine}.

\subsubsection{OpenPIV}

The PIV raw images acquired at the wake downstream of a body are imported into Matlab GUI using \texttt{File->Load}. The user defines the main PIV parameters, including: magnification or scale (i.e.: pixels/meter), time interval between laser pulses, ($\Delta t$), image pre-processing function (e.g. contrast enhancement or inversion of shadowgraphy images), ROI (region-of-interest) for the PIV analysis, interrogation window size and spacing/overlap size (pixels), and filters: signal-to-noise ratio (S/N) type and threshold and an outlier velocity threshold. 


The OpenPIV utilizes a cross-correlation algorithm that yields a displacement vector map by correlating two consecutive PIV images. The output data in pixels of displacement is converted to physical units (meters/second) using the $\Delta t$ and magnification (scale).

\newpage
The main \texttt{openpivgui.m} loop consists of the following steps: 
\begin{itemize}
    \item Cross-correlation algorithm is applied to sub-image square or rectangular interrogation windows. OpenPIV utilizes an FFT-based cross-correlation algorithm to process pairs of PIV images to yield the velocity field maps~\cite{raffel2018}, therefore sizes are typically of $2^n \times 2^n$ size ($32 \times 32, 16 \times 64$, etc.). The spacing/overlap value controls the spatial resolution of the grid $x,y$ at which we estimate horizontal and vertical velocity components ($u, v$). Larger interrogation window size reduces resolution but less affected by the background noise. 
    
    \item The ratio of the maximum cross-correlation peak to the average cross-correlation value, or the ratio of the peak to the second highest peak are used as a measure of the signal-to-noise ratio (S/N)~\cite{Huang1997,Huang1993a,Huang1993b}. Selection of S/N type is based on the data quality (i.e., strong contrast images). The choice of S/N threshold value that marks erroneous vector is obtained through a trial and error procedure, based on manual assessment of the vector field.

    \item After the cross-correlation, additional filters are applied for validation and removal of outliers marked by the S/N ratio or removed based on statistics of the flow field. A so-called global filter removes vectors with length that are larger than the mean of the flow field plus $N$ times its standard deviation. The outlier filter parameter in the GUI indicates the value of $N$. 
    
    \item Local filter is performed on small neighborhoods of vectors using the $3 \times 3$ kernel, removing vectors that are more than 3 times local standard deviation distant from the local mean of the 8 nearest neighbor vectors. 
    
    \item It is desired to complete the analysis with less than 5-10\% of erroneous vectors. After removing the outliers, the missing values are filled using iterative interpolation, based on the valid neighborhood vectors.
\end{itemize}

The velocity field result is stored in three ASCII files: the raw vector results (\texttt{dataName\_noflt.txt}), filtered results (\texttt{dataName\_flt.txt}) and interpolated data (\texttt{dataName.vec}).
Output files have headers that define the list of variables, the units (e.g. pixels/dt or m/s) and the size of the field in terms of rows and columns, followed by the 5 columns of data: $x,y,u,v$, and S/N.

\subsubsection{Spatial and Temporal Analysis Toolbox}

The Spatial and Temporal Analysis Toolbox GUI (\texttt{spatialbox.m}) loads series of \texttt{dataName.vec} files into a 3D flow velocity array where the 3rd dimension is the number of the flow field in the ensemble (or time for the time-resolved case). The data is automatically decomposed (using the Reynolds decomposition) into mean and turbulent fluctuations and provides both qualitative and quantitative visualization tools of a large variety of flow properties. Qualitatively, the toolbox offers several options to show the calculated flow properties in the form of colored contour maps, colored contour lines and vector representation. Quantitatively, it offers the user to plot the properties in a 2D profile format, by selecting regions of interests, or cross-sections. The flow properties include the spatial velocity derivatives and the properties such as vorticity, rate of strain, along with the turbulent parameters such as turbulent intensity, Reynolds stress, turbulent kinetic energy, production, dissipation and enstrophy. Furthermore, the toolbox includes an additional feature allowing to perform some basic spectral analysis using auto-correlation functions applied to the velocity field. The flow characteristics computed in this toolbox for the velocity maps are then exported as a binary Matlab MAT file.


For more details, the reader is referred to the following illustrative example and documentation listed in Table~\ref{table1}.

\subsubsection{getWAKE Toolbox}\label{getWAKEfunctions}

The getWAKE toolbox is designed for wake data analysis, and specifically for the case where several consecutive flow velocity maps contain time evolution of the flow in the wake or motion of the same vortices as they shed behind the body. The toolbox allows to define multiple experimental setup related parameters, required for the reconstruction (i.e. combination) of the wake signature behind a body, followed by the forces estimate procedure. 
The parameters include the PIV parameters, body parameters (dimensions and weight) and an incoming flow parameters (free stream velocity, fluid density and dynamic viscosity) in SI units and the motion type (stationary or cyclic flapping motion, etc.). 
For the flapping case, for instance, the parameters relate to three phases of downstroke, transition and upstroke. 
In some cases, a sub-region from the velocity map is selected mainly to remove noise at the edges. 
For the wake reconstruction, velocity maps sequence parameters and the cross-correlation flow parameter (different then the one used in the OpenPIV analysis, applied to the instantaneous or fluctuating velocity fields) are defined. Further details regarding the wake reconstruction scheme are available in \ref{app1}.


The reconstructed wake signature is presented using the velocity fluctuations vector fields ($u^\prime$, $v^\prime$) and colored with the normalized spanwise vorticity field $\omega_z c/U_\infty$, where $\omega_z=\partial v/\partial x - \partial u/\partial y$, $c$ is the characteristic length scale of the body and $U_\infty$ is the freestream velocity. Multiple visualization options are available, including contour threshold, vorticity or swirl strength~\cite{Zhou1999} contours, with or without Gaussian smoothing, etc. The reconstructed wake can be shown and exported as a static image or as an animated movie.  

The wake data enables the estimation of the profile drag and cumulative circulatory lift coefficients. The lift is calculated using either i) Panda and Zaman method~\cite{Panda1994} or ii) a direct summation of the circulation values. Further details regarding the forces estimation are given in \ref{app2}.

Forces can be presented as a function of the normalized streamwise wake distance ($x/c$, where $c$ is the body size, or chord length of the wing) or time ($t$). If the wake is generated due to a flapping wing cyclic motion, one can correlate the wingbeat phases (downstroke, upstroke) with the velocity maps in the wake by setting the flapping wingbeats kinematics inputs. 
Thus, the force coefficients can be presented for a specific wingbeat or a sequence of wingbeats that correspond to the reconstructed wake signature, where the different wingbeat phases are highlighted.

\section{Illustrative Example}
\label{sec:illustrative_examples}

\begin{figure}[ht!]
	\centering
	\includegraphics[width=\textwidth]{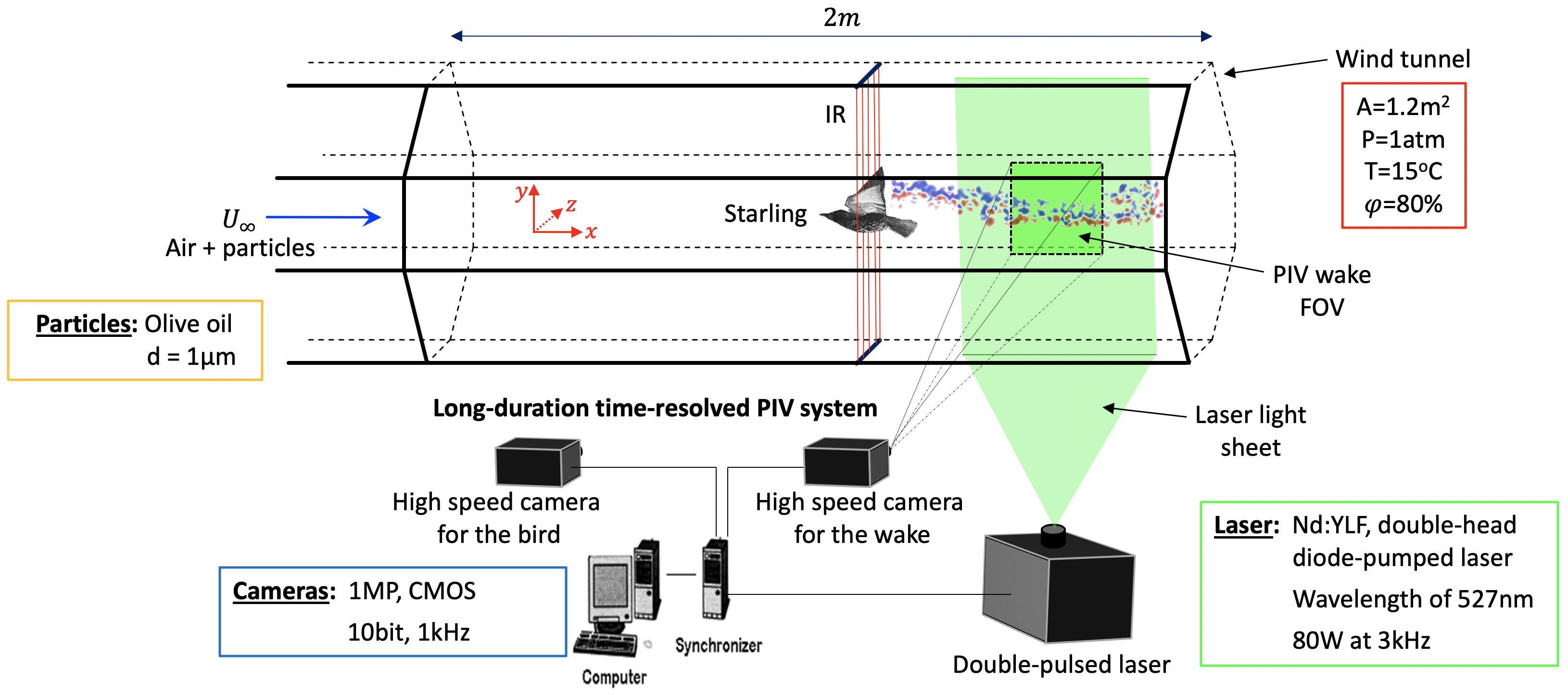}
	\caption{Illustrative scheme of the avian wind tunnel and the experimental setup system. Reproduced from~\cite{Ben-Gida2013}.}
	\label{fig:exp_setup}
\end{figure}

We exemplify the analysis of PIV wake data measurements obtained in the wake of a freely flying European starling (\emph{Sturnus vulgaris}) in a wind tunnel; see Kirchhefer et al.~\cite{Kirchhefer2013} and Ben-Gida et al.~\cite{Ben-Gida2013} for details.
The experimental setup utilized for PIV measurements behind the bird is depicted in \Cref{fig:exp_setup}.  

Experiments were conducted in a climatic closed-loop wind tunnel at the Advanced Facility for Avian Research (AFAR), at Western University (Canada). The wind tunnel test section with cross-sectional area of $1.2\mathrm{m^2}$ is specifically designed for simulating the flight conditions experienced by birds during long distance migratory travel. The flight conditions reported in this work correspond to atmospheric static pressure, a temperature of $15^{\circ}\mathrm{C}$, and relative humidity of $80\%$.

A European starling (\emph{Sturnus vulgaris}) had been trained to fly in flapping flight mode in the wind tunnel. At the time the experiments were performed the bird had a mass of $76\mathrm{g}$, where its wings had an average chord of $c=6\mathrm{cm}$, a maximum wingspan of $b=38\mathrm{cm}$ and an aspect ratio (wingspan squared divided by the wings lifting area) of $\mathrm{AR}=6.4$. A typical cruising flight speed of $U_\infty=13.5\mathrm{m/sec}$ was chosen for the experiments~\cite{Kirchhefer2013}. 

Flow measurements were taken using a long-duration time-resolved PIV system, developed by Taylor et al.~\cite{Taylor2010}, consisting of a $80\mathrm{W}$ double-head diode-pumped Q-switched Nd:YLF laser at a wavelength of $527\mathrm{nm}$ and two CMOS cameras (Photron FASTCAM-1024PCI) with a spatial resolution of $1\mathrm{MP}$ at a sampling rate of $1\mathrm{kHz}$ (see \Cref{fig:exp_setup}).
Olive oil aerosol particles with an average size of $1\mathrm{\mu m}$ \cite{Echols1963} were introduced into the wind tunnel using a Laskin nozzle from the downstream end of the test section. 
One camera was used for measuring the PIV wake data (in the streamwise-normal plane) whilst a second one was used for recording the bird kinematics simultaneously with the PIV, where the PIV wake measurements were taken 2 chord lengths behind the starling. The fields of view of the PIV the bird kinematics cameras were $2c\times2c$ and $9c\times9c$, respectively.
The PIV system sampled pairs of wake images at a rate of $500\mathrm{Hz}$ ($2\mathrm{msec}$ intervals), which allows sufficient resolution for temporally resolving the wake of the bird that flap its wings at a frequency of $15\mathrm{Hz}$.

The distance between the bird wing and the location of the laser sheet was determined by placing a set of IR detectors that trigger the laser once the bird passed the laser location about 3-4 chord length upstream (see \Cref{fig:exp_setup}). 

\subsection{OpenPIV analysis}

\begin{figure}[ht!]
	\centering
	\begin{subfigure}[b]{0.315\textwidth}
		\centering
		\includegraphics[width=\textwidth]{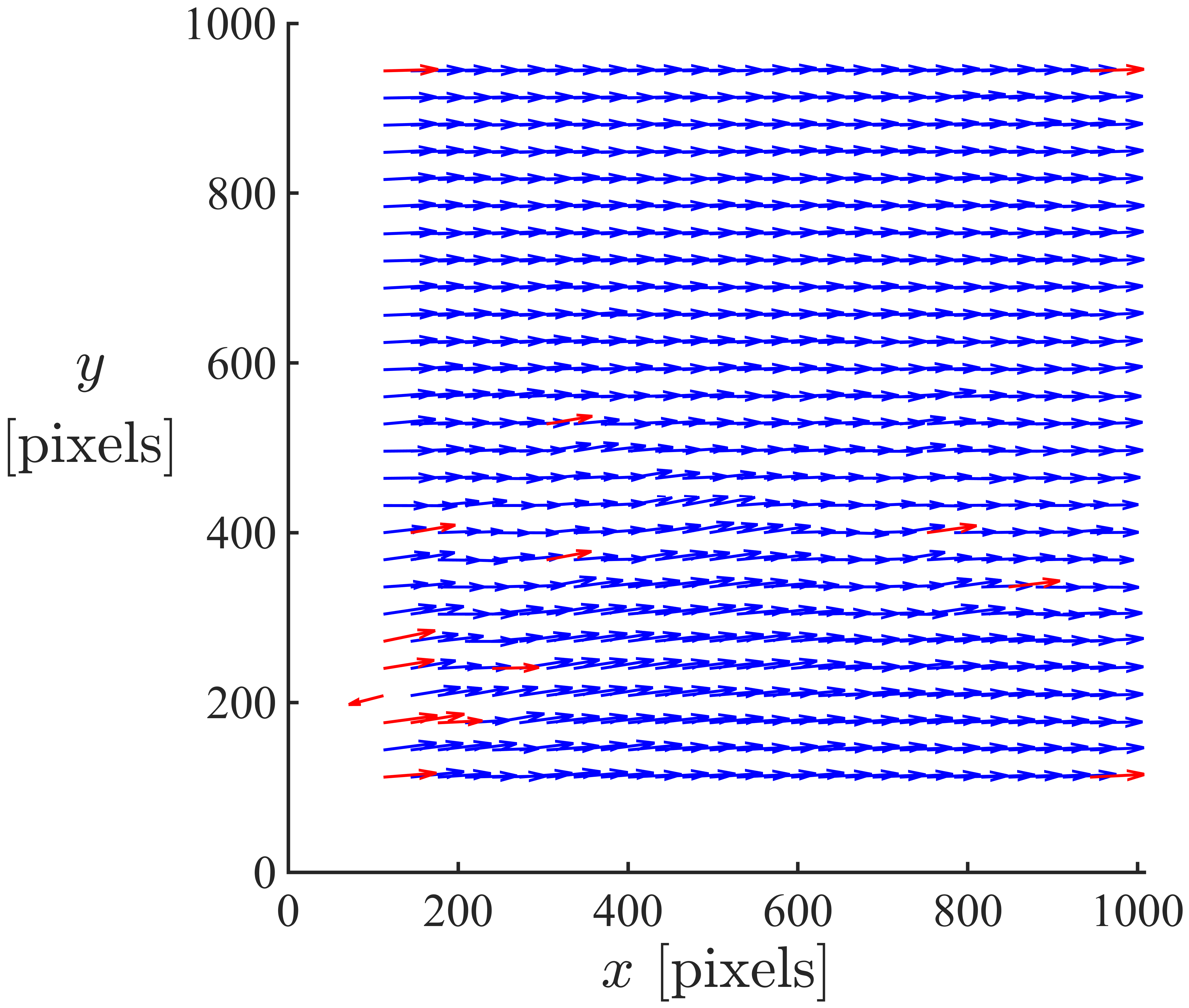}
		\caption{}
		\label{fig:noflt_vec_file}
	\end{subfigure}
~
	\begin{subfigure}[b]{0.315\textwidth}
		\centering
		\includegraphics[width=\textwidth]{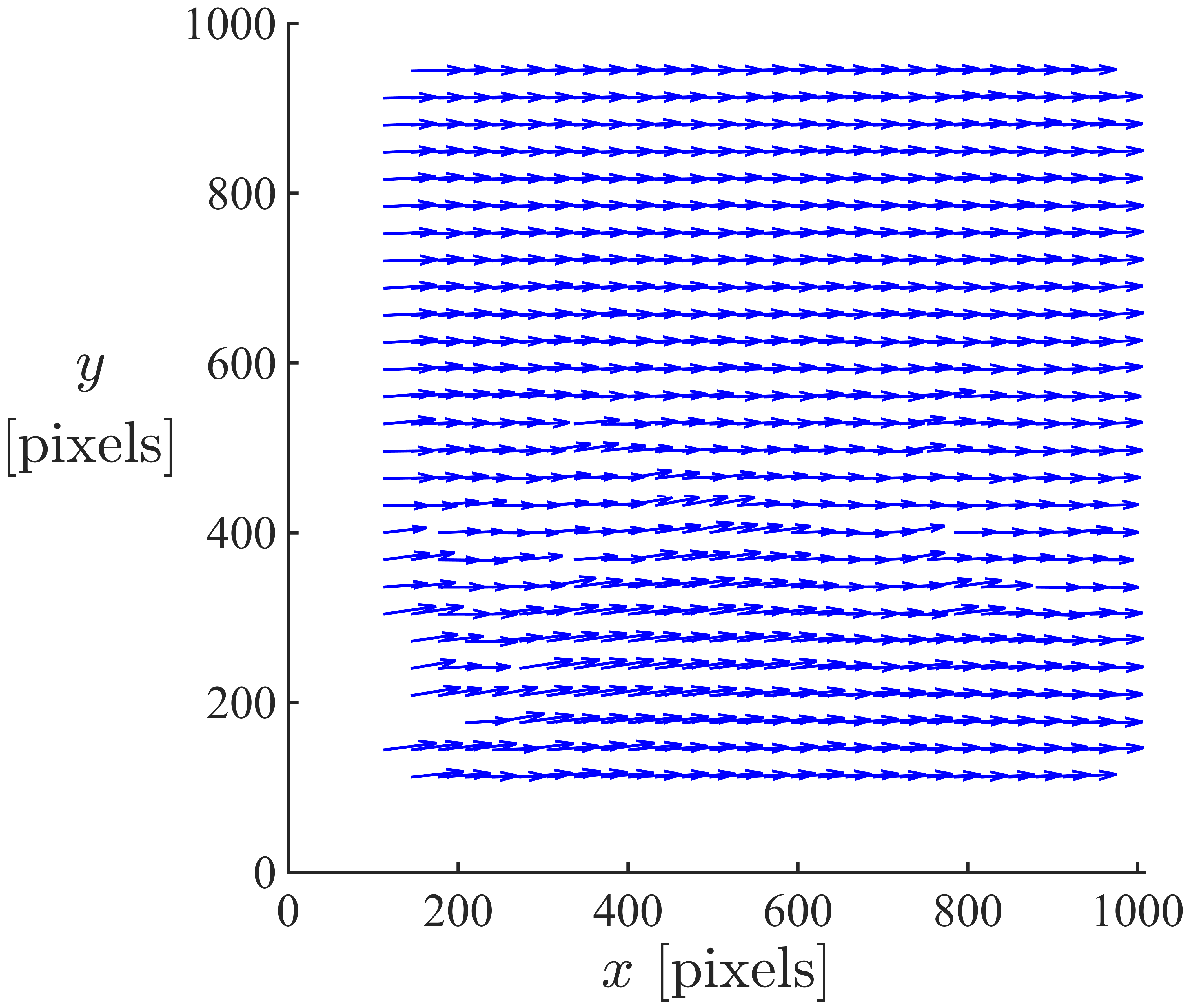}
		\caption{}
		\label{fig:flt_vec_file}
	\end{subfigure}
~
	\begin{subfigure}[b]{0.315\textwidth}
		\centering
		\includegraphics[width=\textwidth]{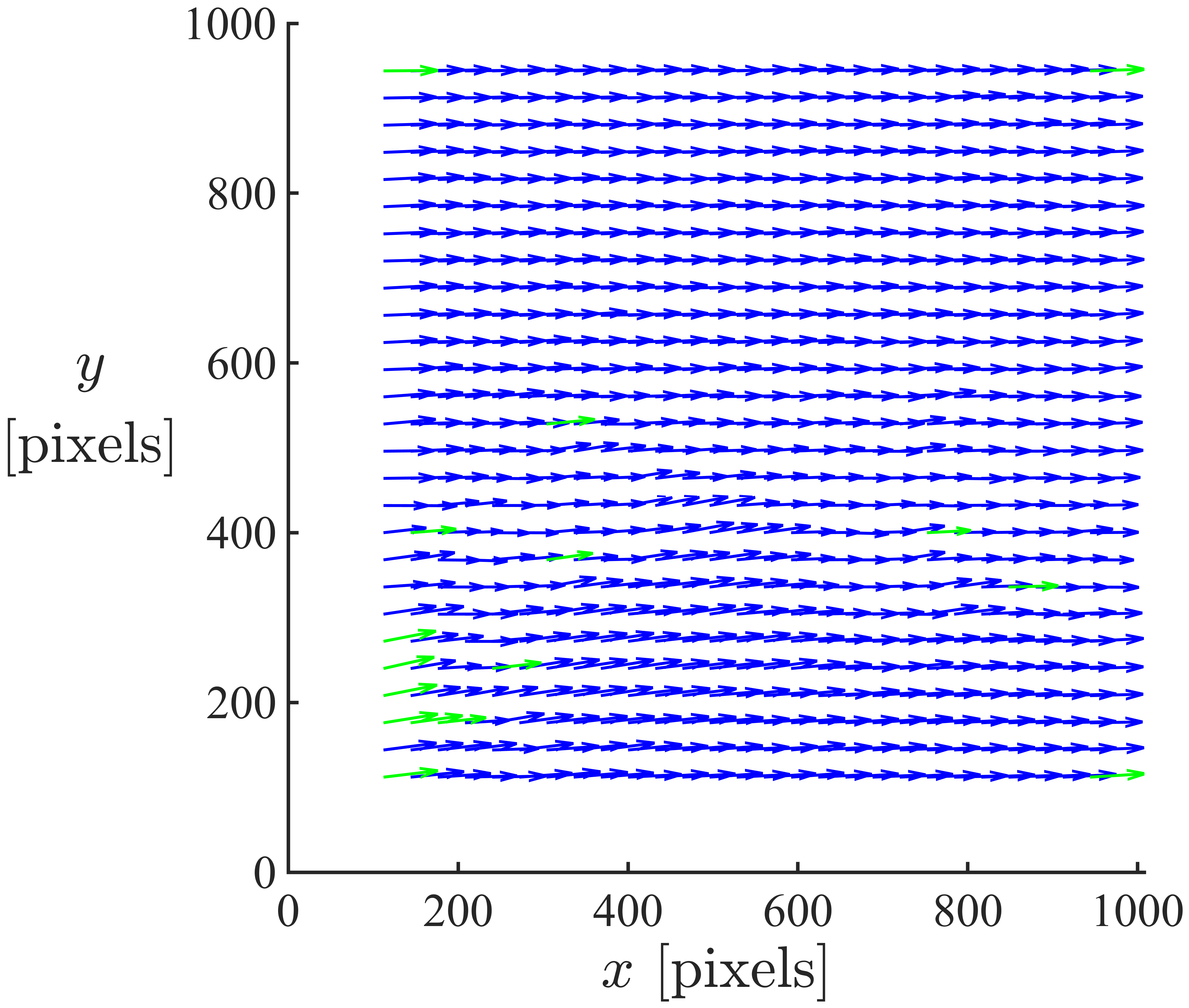}
		\caption{}
		\label{fig:final_vec_file}
	\end{subfigure}
	\caption{An example of the output velocity vector maps files from the OpenPIV analysis in Matlab: (a) raw, (b) filtered, and (c) interpolated velocity vector maps, respectively. Outlier velocity vectors are colored in red, whereas interpolated vectors are colored in green.}
	\label{fig:OpenPIV_vec_outputs}
\end{figure}

The velocity fields were computed from the PIV wake data behind the bird with $32\times32\mathrm{pixel^2}$ interrogation windows and 50\% overlap ($16\times16\mathrm{pixel^2}$), thus yielding a spatial resolution of 32 vectors per average chord of the bird ($c$), equal to 1.8 vectors per millimeter. Here, the second type of the S/N type was used (with a threshold value of 1) and an outlier filter value of 100 was chosen. An example of velocity vector maps computed by the OpenPIV is depicted in \Cref{fig:OpenPIV_vec_outputs}. 
The raw velocity map (non-filtered) is depicted in \Cref{fig:noflt_vec_file}, where the outlier vectors are colored in red. The filtered velocity map (after applying global and local filters) is depicted in \Cref{fig:flt_vec_file}. 
The final velocity map is shown in \Cref{fig:final_vec_file}, where the interpolated vectors are colored in green.

\begin{figure}[ht!]
	\centering
	\begin{subfigure}[b]{0.63\textwidth}
		\centering
		\includegraphics[width=\textwidth]{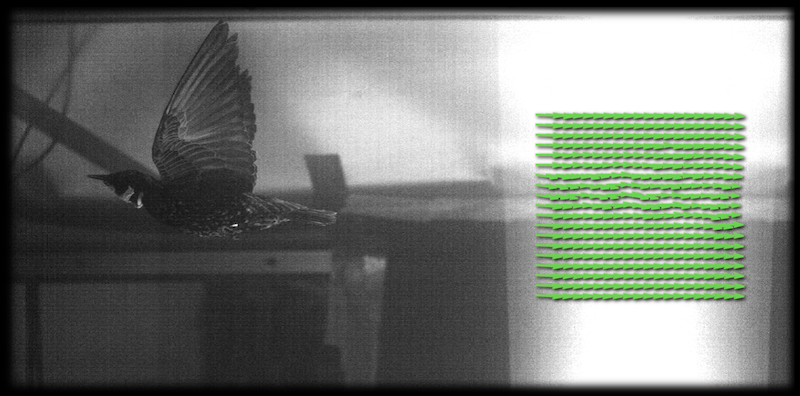}
		\caption{}
		\label{fig:openPIV:velocity_map_downstroke}
	\end{subfigure}
	
	\begin{subfigure}[b]{0.63\textwidth}
		\centering
		\includegraphics[width=\textwidth]{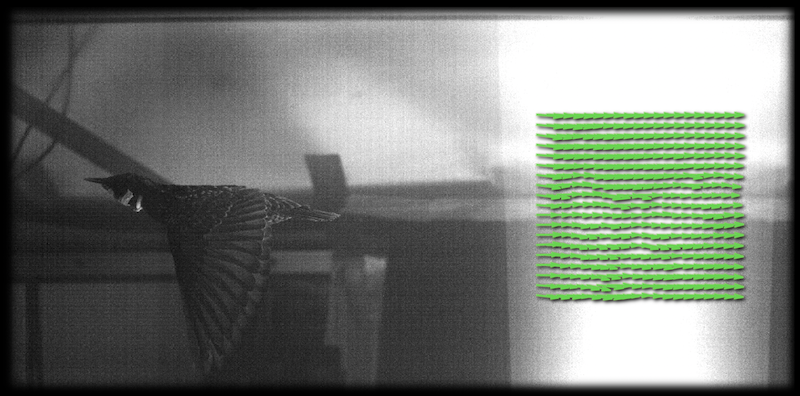}
		\caption{}
		\label{fig:openPIV:velocity_map_upstroke}
	\end{subfigure}
	\caption{Example instantaneous velocity vector maps, as computed from the OpenPIV (\texttt{opevpivgui.m}) at the near wake behind the flapping starling, while it initiates (a) the downstroke phase and (b) the upstroke phase.}
	\label{fig:openPIV:velocity_map_example}
\end{figure}

\Cref{fig:openPIV:velocity_map_example} depicts an example of instantaneous velocity maps computed using the OpenPIV (\texttt{opevpivgui.m}) at the near wake behind the starling, while it initiates the downstroke and upstroke phases.
The bright light on the right side of the image is the laser light sheet illuminating the flow field behind the bird.

\begin{figure}[ht!]
	\centering
	\begin{subfigure}[b]{0.42\textwidth}
		\centering
		\includegraphics[width=\textwidth]{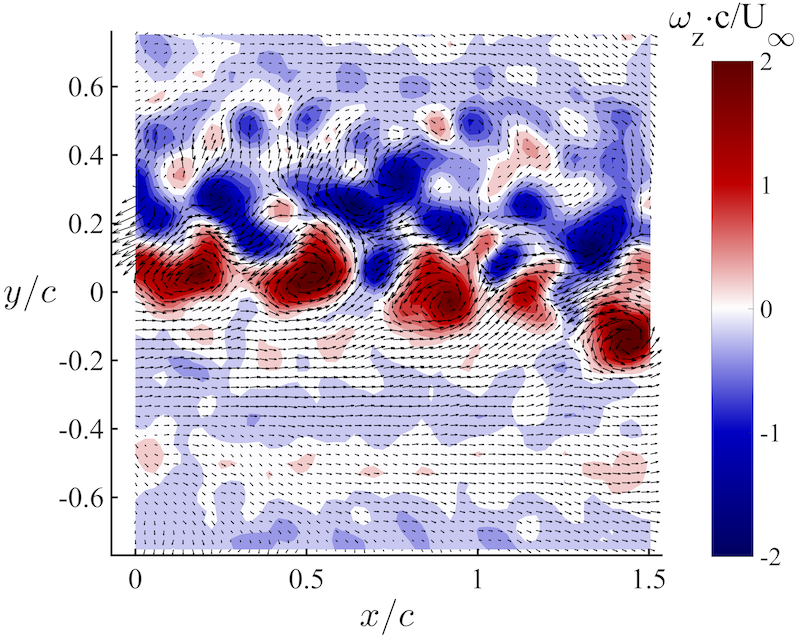}
		\caption{}
		\label{fig:spatial:vorticity_map_downstroke}
	\end{subfigure}
	~
	\begin{subfigure}[b]{0.42\textwidth}
		\centering
		\includegraphics[width=\textwidth]{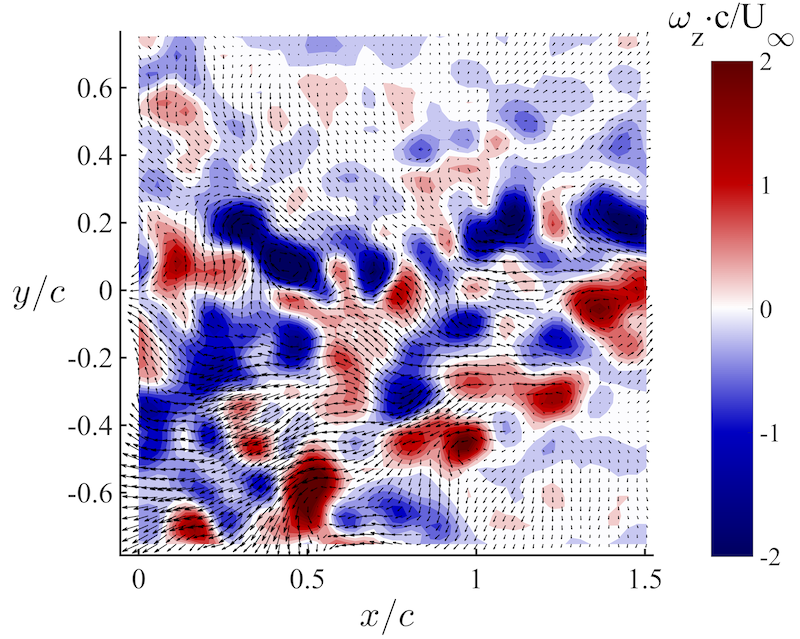}
		\caption{}
		\label{fig:spatial:vorticity_map_upstroke}
	\end{subfigure}
	\caption{Example instantaneous spanwise vorticity contour fields, as computed from the Spatial and Temporal Toolbox (\texttt{spatialbox.m}) at the near wake behind the flapping starling, while it initiates (a) the downstroke phase and (b) the upstroke phase. The air flows from left to right and the vectors displayed are the velocity fluctuations.}
	\label{fig:spatial:vorticity_map_example}
\end{figure}

The final velocity vector maps in the wake, which were computed using the OpenPIV (\texttt{opevpivgui.m}) and exported as .vec files, are then imported into the Spatial and Temporal Analysis Toolbox (\texttt{spatialbox.m}) for the computation of the required flow characteristics.
\Cref{fig:spatial:vorticity_map_example} depicts an example of the instantaneous spanwise vorticity fields computed using the Spatial and Temporal Analysis Toolbox, from the velocity vector maps depicted in \Cref{fig:openPIV:velocity_map_example}, in the near wake behind the starling during flight.

\newpage

\begin{figure}[ht!]
	\centering
	\includegraphics[width=\textwidth]{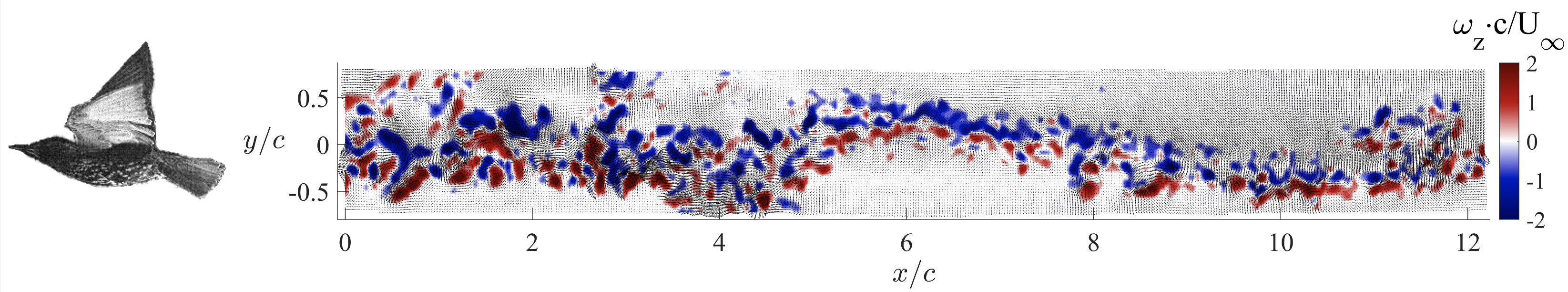}
	\caption{Example of a wake reconstructed by the getWAKE Toolbox from PIV wake measurements taken behind a freely flying starling during a complete flapping wingbeat. The contour is the normalized spanwise vorticity in the wake, $\omega_z c/U_\infty$. The bird flew from right to left; therefore, the downstream distance is measured as positive chord lengths, $x/c$. What appears as downstream essentially happened earlier, while what appears as upstream happened later. The vectors displayed are the velocity fluctuations. A threshold of 8\% from the global maximum absolute vorticity values in the wake field was applied.}
	\label{fig:GUI-wake-evolution-example}
\end{figure}

The flow characteristics and the velocity vector maps are then exported from the Spatial and Temporal Analysis Toolbox as a binary MAT file, and imported into the getWAKE Toolbox GUI (\texttt{wake.m}). The characteristics of the experimental setup (PIV system, bird and freestream flow) are set in the getWAKE GUI. Using these inputs, the wake signature behind the starling is reconstructed for a sequence of velocity maps, which corresponded to flapping wingbeat phases: downstroke and upstroke, as depicted in \Cref{fig:GUI-wake-evolution-example}. The bird appears to fly from right to left; thus, the downstream wake essentially occurred earlier (corresponding to the downstroke phase), while the upstream wake occurred later (corresponding to the upstroke phase). Here, we utilized the velocity fluctuations ($u^{\prime},v^{\prime}$) for cross-correlating the velocity maps.

\begin{figure}[ht!]
	\centering
	\begin{subfigure}[b]{0.48\textwidth}
		\centering
		\includegraphics[width=\textwidth]{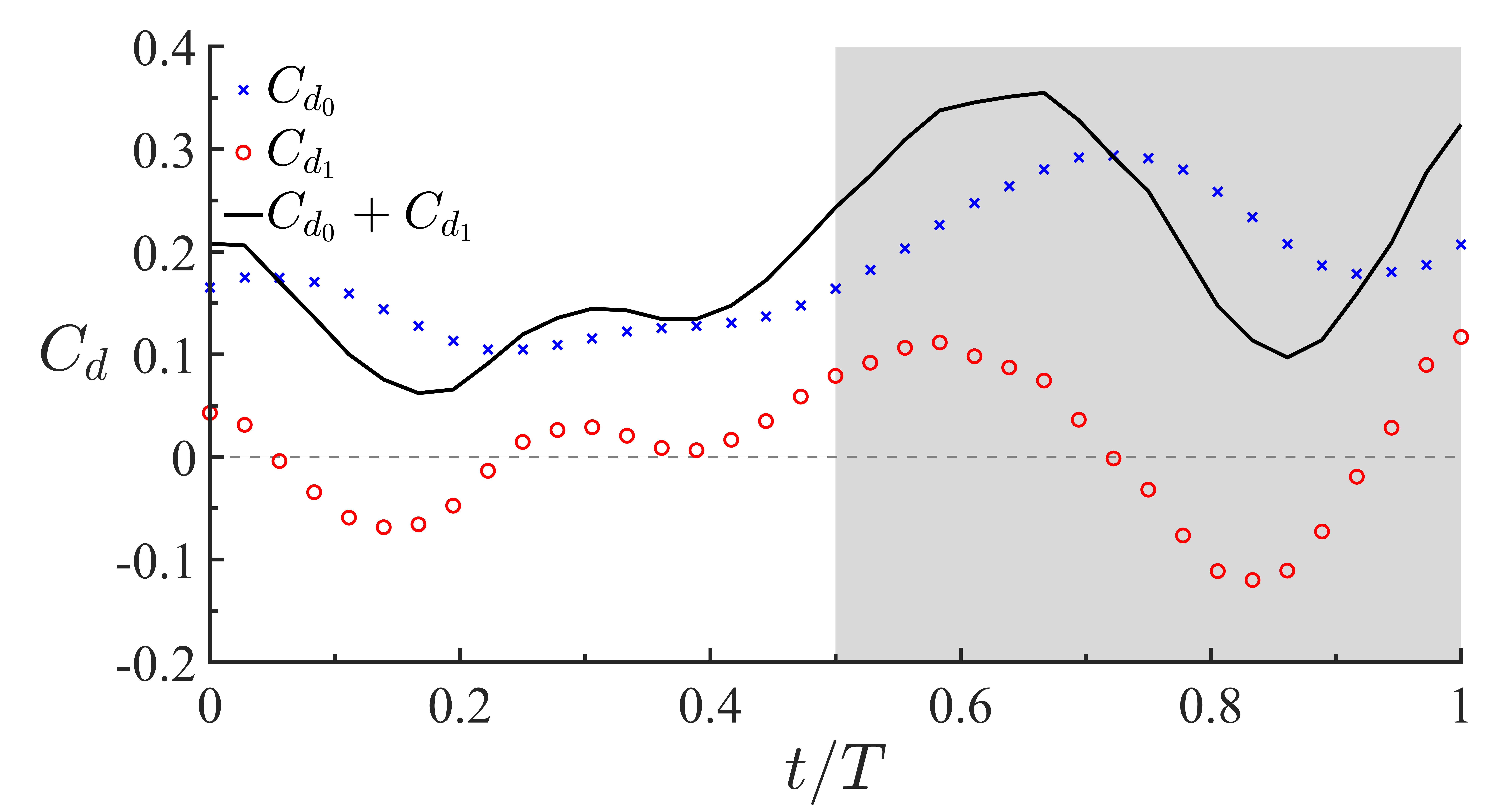}
		\caption{}
		\label{fig:getWake:drag_vs_time}
	\end{subfigure}
	~
	\begin{subfigure}[b]{0.48\textwidth}
		\centering
		\includegraphics[width=\textwidth]{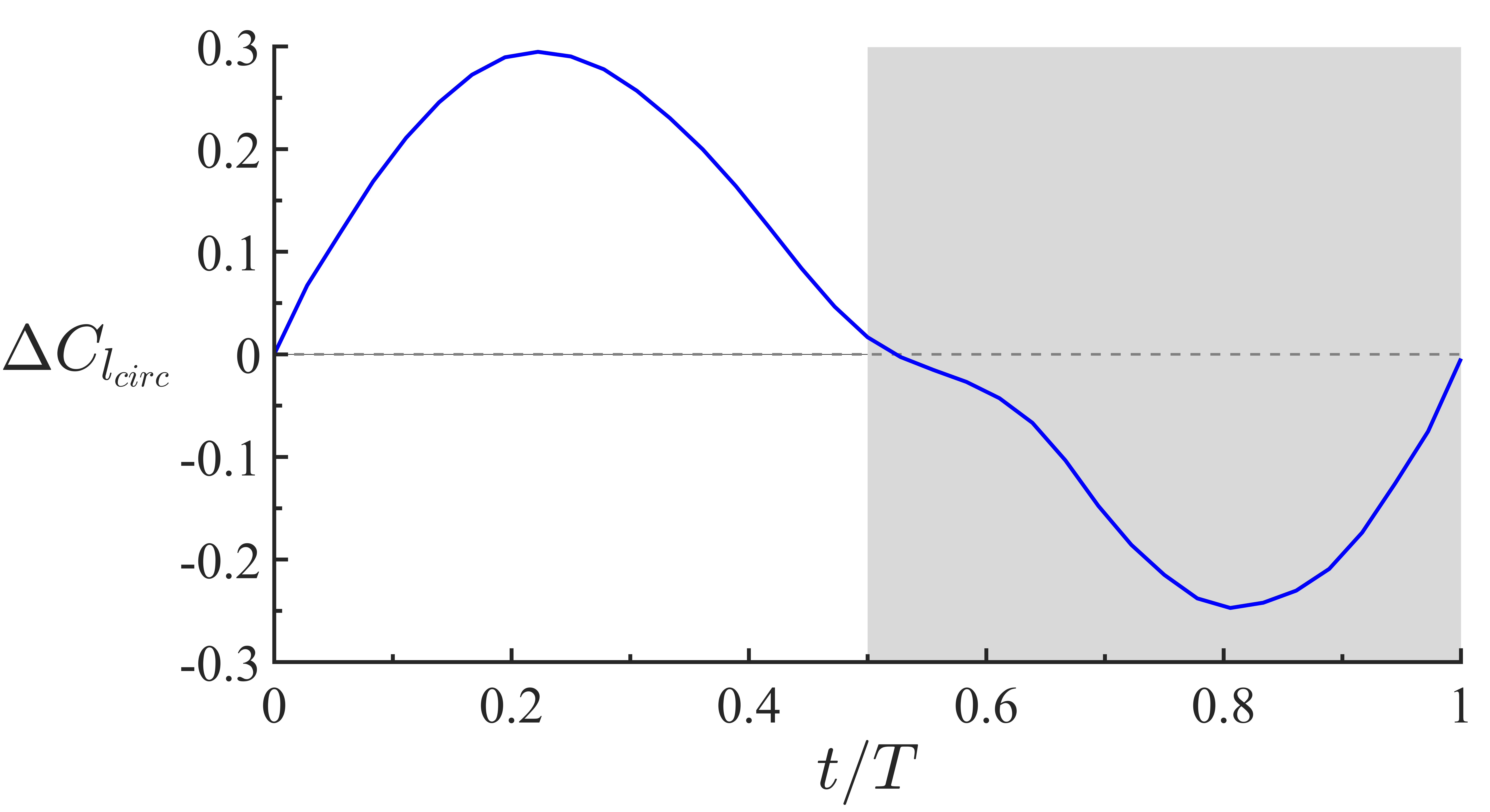}
		\caption{}
		\label{fig:getWake:lift_vs_time}
	\end{subfigure}
	
	\centering
	\begin{subfigure}[b]{0.48\textwidth}
		\centering
		\includegraphics[width=\textwidth]{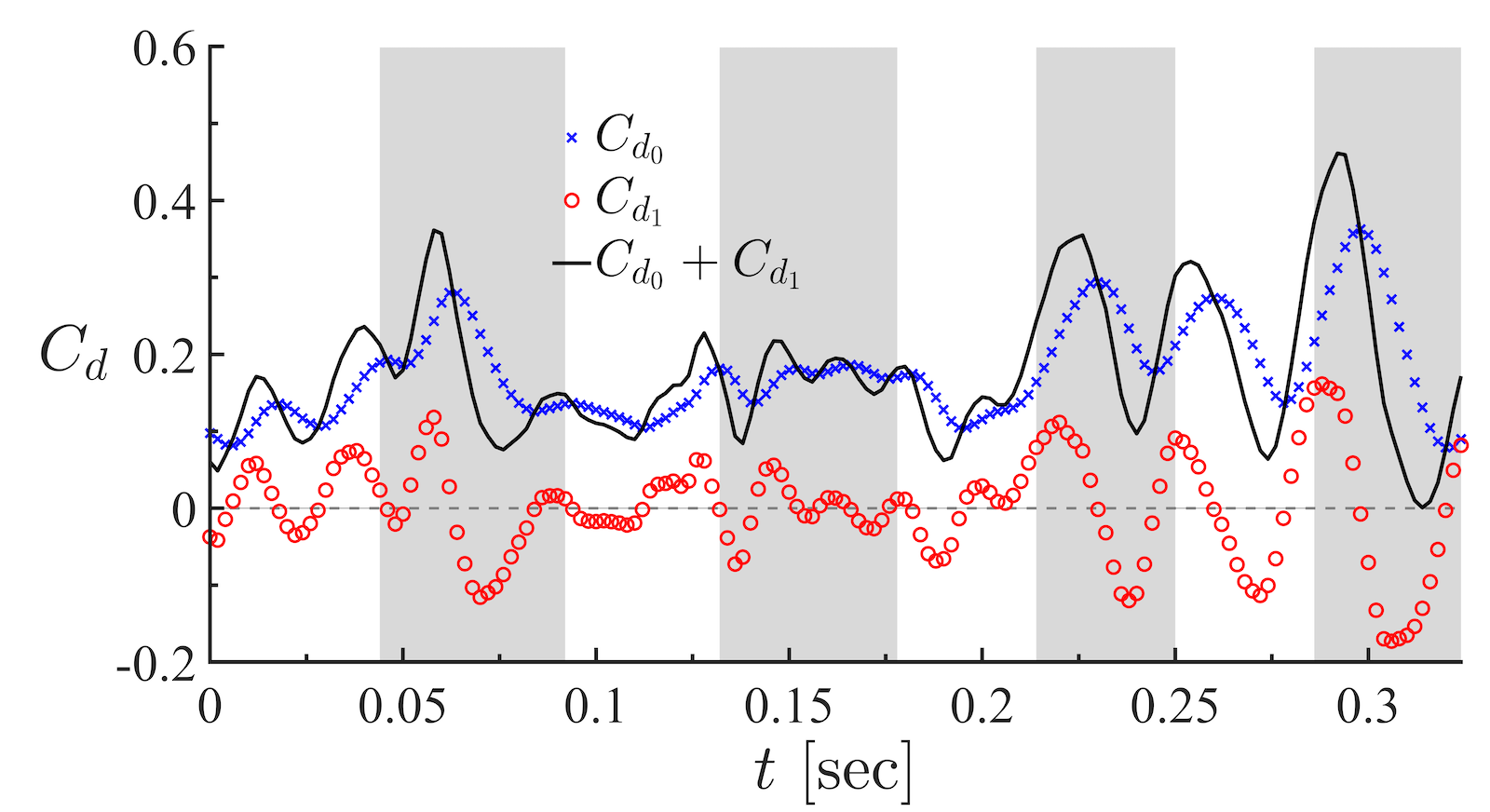}
		\caption{}
		\label{fig:getWake:drag_vs_time_4wingbeats}
	\end{subfigure}
	~
	\begin{subfigure}[b]{0.48\textwidth}
		\centering
		\includegraphics[width=\textwidth]{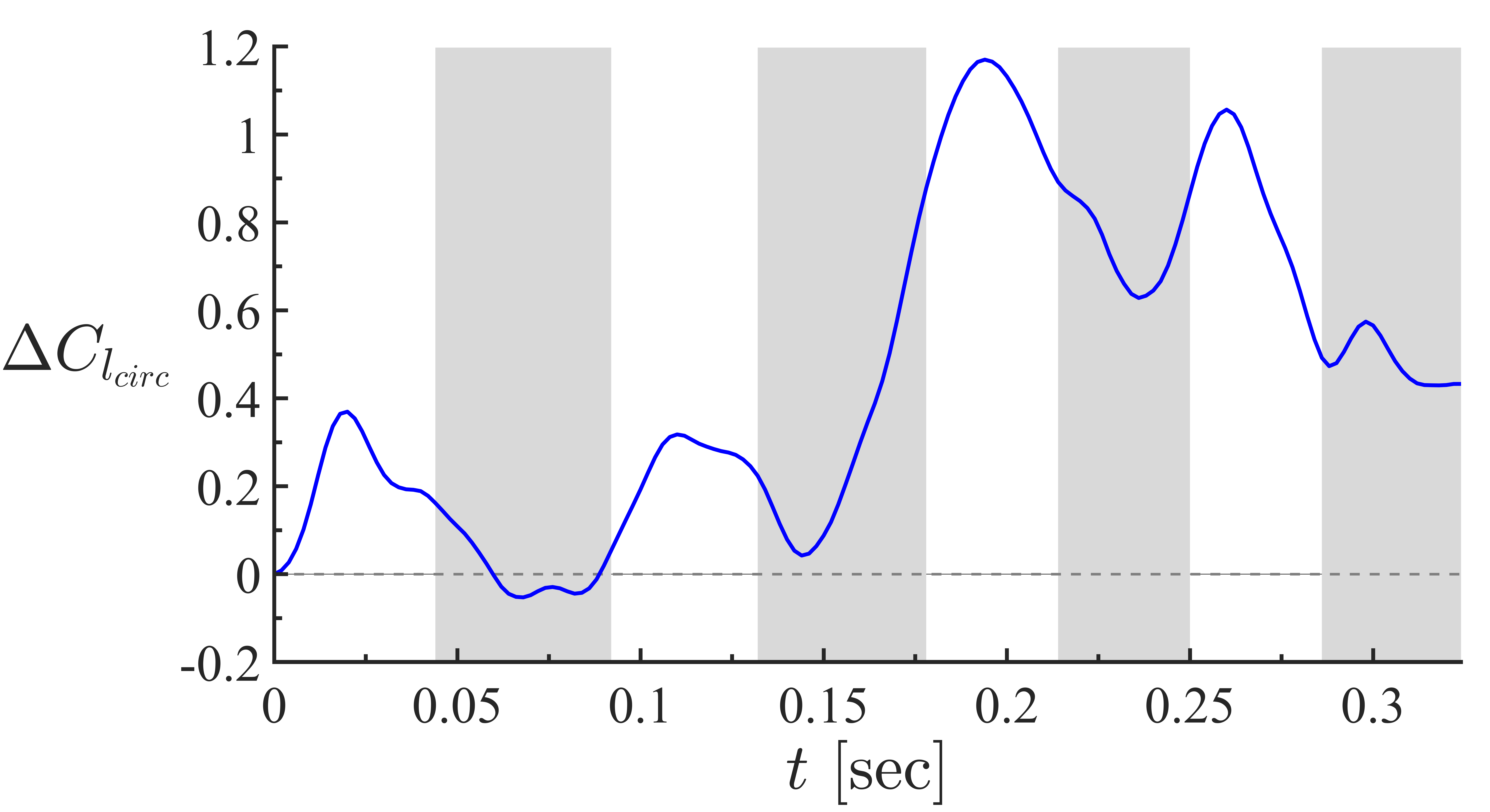}
		\caption{}
		\label{fig:getWake:lift_vs_time_4wingbeats}
	\end{subfigure}
	\caption{Time variation of the drag coefficient $C_d$ and the cumulative circulatory lift coefficient $\Delta C_{l_{circ}}$ for (a-b) a single flapping wingbeat, and (c-d) four consecutive flapping wingbeats of the starling. The white shaded regions correspond to the downstroke phases, whereas the gray shaded regions correspond to the upstroke phases.}
	\label{fig:getWake:forces_vs_time}
\end{figure}

The time variation of the profile drag and cumulative circulatory lift forces exerted on the bird is depicted in \Cref{fig:getWake:forces_vs_time}. We demonstrate the forces calculations over a single and multiple wingbeat phases.   
The time variation of the forces for a single wingbeat are given as a function of the non-dimensional time, $t/T$; where $T$ is the time period of the flapping wingbeat cycle. 
The white shaded regions in \Cref{fig:getWake:forces_vs_time} correspond to the downstroke phases of each flapping wingbeat, whereas the gray shaded region corresponds to the upstroke phases. 
Herein, $C_d$, total drag coefficient is the summation of the $C_{d_0}$ and $C_{d_1}$ components, which are the steady and unsteady components of the drag coefficient, respectively. 
The cumulative circulatory lift coefficient was computed based on Panda and Zaman method \cite{Panda1994}, see the procedure in  \ref{app2}).

\section{Impact}
\label{sec:impact}

The main contribution of this work is to provide an open source code for PIV data analysis and post-analysis. OpenPIV-Matlab is a complete 2D PIV data analysis software package that can handle PIV and its extensions (i.e., time-resolved, stereo) images and perform a detailed flow analysis.
To the best of our knowledge, OpenPIV-Matlab is the most complete set of tools for the flow analysis based on data acquired using optical measurement techniques in fluid dynamics, providing thousands of lines of code as an open source. The toolbox is used in a variety of fields associated with fluid dynamics such as such as mechanical, aerospace, civil and environmental engineering, biology, chemistry, earth sciences and so forth.
The package presented herein offers the only known open-source software for advanced post-processing of time-resolved PIV wake data behind immersed bodies, enabling visual description of the wake evolution over time and space estimation of the aero/hydrodynamic forces.

\section{Conclusions}
\label{sec:conclusions}
We present a Matlab (Mathworks Inc.) software package for analysis and post-analysis of PIV data. OpenPIV is comprised of three toolboxes that can analyse PIV images to yield a 2D2C (two-dimensional, two-component) velocity vector fields using cross-correlation technique, post-analysis the velocity fields calculating various flow properties including turbulence and spectral analysis and estimation of the aero/hydrodynamic forces exerted over immersed bodies in fluids. As test-case study of the toolbox applicability to PIV time-resolved data, we present flow analysis of the near wake region behind a freely flying bird in a closed-loop wind tunnel. We demonstrate the functionality and usefulness of the toolboxes as applied to a data collected in a wake region that is unsteady and turbulent. The main contribution is a free, flexible, extendable and validated platform for both basic and advanced PIV related analysis and post-processing.

\section{Conflict of Interest}
We wish to confirm that there are no conflicts of interest associated with this publication and there has been no significant financial support for this work that could have influenced its outcome.

\section*{Acknowledgements}
\label{sec:acknowledgements}

The authors want to acknowledge all the co-authors of the OpenPIV-Matlab toolboxes, listed on the respective Github pages, as listed in Table~\ref{table1}. The open source software could not exist without the vibrant community of users and developers that contribute by their test cases, verification and validation studies, and of course software and documentation development. 



\newpage
\appendix
\setcounter{equation}{0}
\setcounter{figure}{0}
\setcounter{subsection}{0}
\setcounter{subsubsection}{0}
\renewcommand{\theequation}{A.\arabic{equation}}
\renewcommand{\thefigure}{A.\arabic{figure}}

\section{Wake Reconstruction}\label{app1}
The wake reconstruction (see an example wake measured behind a flapping starling in \Cref{fig:GUI-wake-evolution-example}) is based on time-resolved PIV images taken from a stationary camera yielding Eulerian observation of the flow field behind an immersed body (either stationary or undergoing flapping wing motion). The wake reconstruction method described herein was originally developed by the first author and later on certain aspects of the methods appeared in \citep{Kirchhefer2012}. Thus, there are similarities in terms of its core principles, however, the utilization and output of the reconstruction scheme detailed below are different.

For the wake reconstruction, we assume the body's position did not change much relative to the measurement plane. 
Based on Taylor's frozen turbulence hypothesis \cite{Taylor1938}, we assume that the turbulent flow remains relatively unchanged as it passes through the measurement plane. This hypothesis implies that there is no significant variation of a spatial velocity distribution over the timescale required for observation, supported by previous studies, e.g. Zaman and Hussain~\cite{Zaman1981}.

\begin{figure}[ht!]
	\centering
	\begin{subfigure}[b]{0.33\textwidth}
		\centering
		\includegraphics[width=\textwidth]{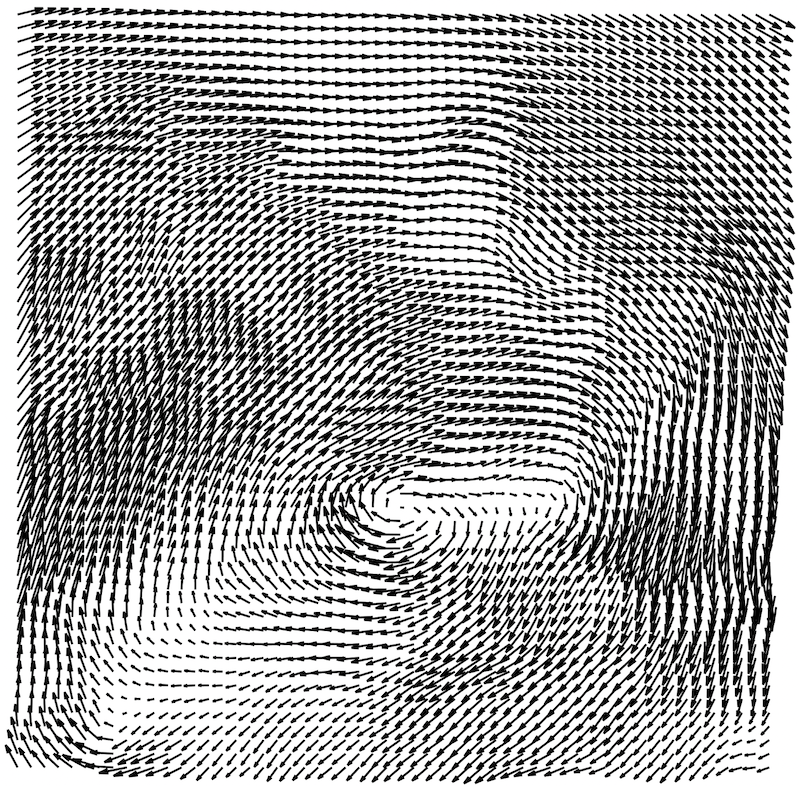}
		\caption{}
		\label{fig:getWAKE:vec_map_t0}
	\end{subfigure}
	~
	\begin{subfigure}[b]{0.33\textwidth}
		\centering
		\includegraphics[width=\textwidth]{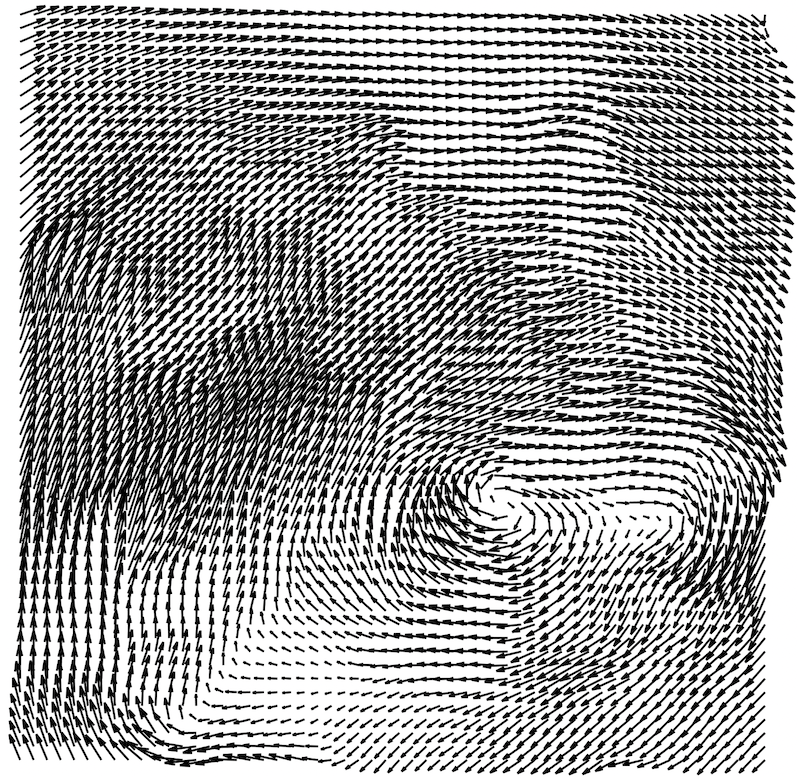}
		\caption{}
		\label{fig:getWAKE:vec_map_t1}
	\end{subfigure}
	
	\begin{subfigure}[b]{0.41\textwidth}
		\centering
		\includegraphics[width=\textwidth]{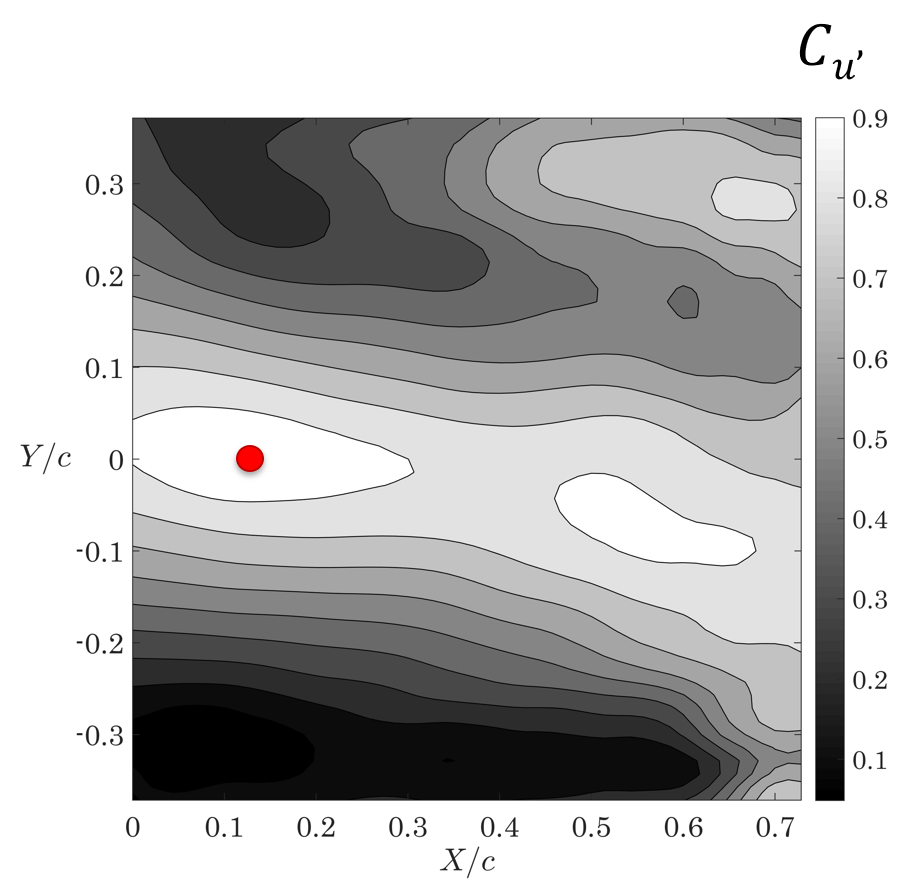}
		\caption{}
		\label{fig:getWAKE:Cu_map}
	\end{subfigure}
	~
	\begin{subfigure}[b]{0.41\textwidth}
		\centering
		\includegraphics[width=\textwidth]{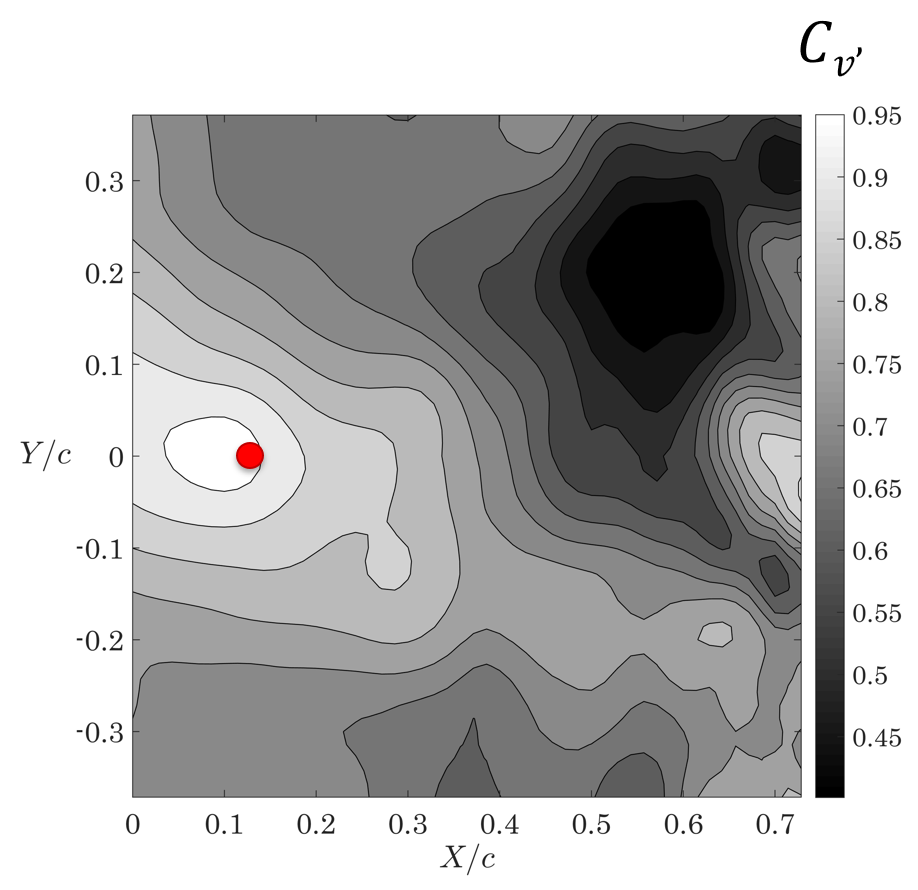}
		\caption{}
		\label{fig:getWAKE:Cv_map}
	\end{subfigure}
	
	\begin{subfigure}[b]{0.45\textwidth}
		\centering
		\includegraphics[width=\textwidth]{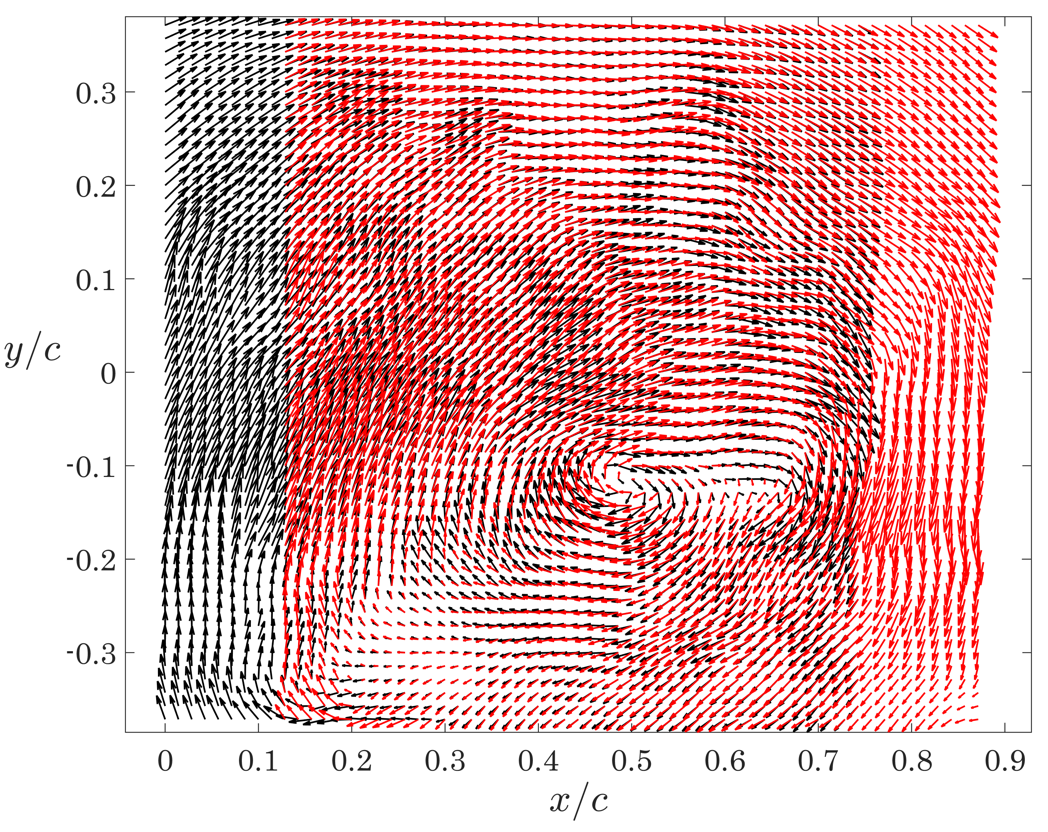}
		\caption{}
		\label{fig:getWAKE:XY_shift_between_maps}
	\end{subfigure}
	\caption{An example of the spatial ($X,Y$) shift, as computed from the cross-correlation procedure, of two consecutive velocity vector maps in the near wake behind a freely flying European starling. The red dots mark the location of the maximum cross-correlation coefficient. (a) velocity vector map at time $t_1$ (the flow is from left to right); (b) consecutive velocity vector map at time $t_2=t_1+2~\mathrm{msec}$; (c) cross-correlation map of the $C_{u^\prime}$ coefficient; (d) cross-correlation map of the $C_{v^\prime}$ coefficient; (e) spatial shift visualization of the two consecutive velocity vector maps.}
	\label{fig:getWAKE:shift_example_Starling_wake}
\end{figure}

As a minimal requirement, sufficient temporal resolution is needed to track the flow patterns as they propagate from one velocity map to another (see \Cref{fig:getWAKE:shift_example_Starling_wake}).
For example, if the wake is measured behind a bird in flapping flight with a flapping frequency of 10$\mathrm{Hz}$, the PIV velocity vector maps must be sampled at a rate of 100$\mathrm{Hz}$ or higher ($\geq200\mathrm{Hz}$ PIV raw images sampling rate); thus, enabling a given flow structure to be tracked by at least three consecutive velocity maps.

The wake composite image is generated by offsetting each consecutive PIV velocity map with a calculated instantaneous convection velocity and then overlap the images, while keeping the mid region of each instantaneous PIV velocity map (to avoid overlapping 'noisy' data from the velocity map edges).
The instantaneous convection velocity, which determines the offset if each PIV velocity map, is calculated based on a cross-correlation algorithm that examine the match of the velocity vector fields ($u,v$) or the fluctuating velocity vector fields ($u^{\prime},v^{\prime}$) of two consecutive velocity maps.
The cross-correlation coefficient, which determines the match of two consecutive PIV velocity maps, is calculated as follows (with respect to a given flow property $s$):
\begin{equation}
\begin{aligned}[b]
&C_s(X,Y,T)=    \\
&\sum_{i=1,j=1}^{I,J} \frac{\left[ s(x_i,y_i,t) - \overline{s(t)} \right] \left[ s(x_i+X,y_i+Y,t+T) - \overline{s(t+T)} \right]}{IJ\sigma_s(t)\sigma_s(t+T)}
\label{eq:Cp}
\end{aligned}
\end{equation}
If $C=1$, the two velocity maps are identical (perfectly matched), whereas if the two images are different, then $C<1$.
($I,J$) is the size of the overlapping area and $\sigma_s$ is the standard deviation of the flow property $s$. Here, $s$ is to be replaced with ($u,v$) for computing the cross-correlation coefficients with respect to the velocity vector components ($C_u,C_v$), or with ($u^{\prime},v^{\prime}$) for computing the cross-correlation coefficients with respect to the velocity fluctuations vector components ($C_{u^{\prime}},C_{v^{\prime}}$).
The spatial shift ($X,Y$) of any instantaneous PIV velocity map is computed based on overlapping that achieves the maximum correlation coefficient; i.e., $\mathrm{max}\left[C_u,C_v\right]$ or $\mathrm{max}\left[C_{u^\prime},C_{v^\prime}\right]$.

If the cross-correlation failed during the wake reconstruction process (e.g., when the PIV data is low in quality or if the computed spatial shift is larger than the velocity map boundaries), the spatial shift of each instantaneous PIV map is determined from the freestream advection velocity $U_\infty$ and the time difference $\Delta t$; i.e., $X=U_\infty\Delta t$.

\newpage

\setcounter{equation}{0}
\setcounter{figure}{0}
\setcounter{subsection}{0}
\setcounter{subsubsection}{0}
\renewcommand{\theequation}{B.\arabic{equation}}
\renewcommand{\thefigure}{B.\arabic{figure}}

\section{Forces Estimation}\label{app2}
The variation of the drag force coefficient $C_d$ can be estimated from the PIV wake data as shown in Ben-Gida et al. \cite{Ben-Gida2013}. The analysis is based on the flow momentum equations that with appropriate assumptions leads to the following formulation of the drag coefficient for the two-dimensional wake case:
\begin{equation}
C_d = \underbrace{\frac{2}{cU_\infty}\int_0^h u\left(1 - \frac{u}{U_\infty}\right) \mathrm{d}y}_{C_{d_0}\text{ - Steady part}} - \underbrace{ \frac{2}{cU_\infty}\frac{\partial}{\partial t}\int_0^h \int_0^l \frac{u}{U_\infty} \mathrm{d}x\mathrm{d}y}_{C_{d_1}\text{ - Unsteady part}}
\label{eq:Cd}
\end{equation}
where ($x,y$) is the Cartesian coordinates system used for the streamwise-normal plane in the wake; the $x$-axis indicates the downstream direction and the $y$-axis is the normal direction.
Conceptually, the steady drag coefficient component $C_{d_0}$ is proportional to the velocity deficit at the wake, whereas the unsteady drag coefficient component $C_{d_1}$ is associated with the unsteady flow motion.
While the steady drag term can be obtained from the near wake velocity field, the unsteady drag term requires information regarding the entire control volume (surface in the case of 2D PIV where we assume volume per unit length) surrounding the body over time. 
We assume most of the unsteady disturbances generated by the unsteady motion are obtained from the velocity field at the near wake where both unsteady contribution and viscous effects have not dissipated yet.
Therefore, we approximate the full surface integral of the unsteady term to include only the velocity field obtained from the PIV experiments in the body near wake. Here, $U_\infty$ is the mean undisturbed streamwise velocity, and $h$ and $l$ are the vertical and horizontal extent of the computed velocity field in the wake, respectively.
For the steady drag coefficient, we average the various profiles along the streamwise extent of each PIV map, thus yielding a single $C_{d_0}$ value to represent each PIV map. 
It is noteworthy that drag coefficient is computed for a sequence of velocity maps, and therefore it does not require \textit{a priori} information of the reconstructed wake.

The variation of the cumulative circulatory lift coefficient $\Delta C_{l_{circ}}$ can be estimated from the PIV velocity vector fields based on Stalnov et al. \cite{Stalnov2015}, Ben-Gida et al. \cite{BenGida2016} and Nafi et al. \cite{Nafi2020}.
Any fluid motion around a body is accompanied by the shedding of vortices into the wake.
By analyzing these vortical patterns in the near wake, one can estimate the unsteady lift exerted on the body. 
Herein, the estimation of increment in the time-dependent lift throughout the wake is evaluated from the PIV velocity fields by utilizing Wu's viscous flow approach \cite{Wu1981}, which was later expressed by Panda and Zaman \cite{Panda1994}.
This method can only be applied for near wake flow measurements behind bodies, where it is assumed that the wake has not deformed yet and interactions between the vortices (which results from the tip and root regions) shed into the wake are not significant. 
Assuming two-dimensional, incompressible flow and neglecting added mass effects, the time-dependent circulatory lift force exerted on a body can be evaluated from the near wake flow field, as follows \cite{Panda1994}:
\begin{equation}
\begin{aligned}[b]
L_{circ}(t) &= \underbrace{\rho \frac{\mathrm{d}}{\mathrm{d}t}\iint_Ax\omega_z(t)\mathrm{d}x\mathrm{d}y}_{x\text{-moment of the vorticity field}} + \\
&\underbrace{\rho U_\infty \int_0^t \int_0^h\nu \left( \frac{\partial^2u(t)}{\partial x^2} + \frac{\partial^2u(t)}{\partial y^2} \right)\mathrm{d}y \mathrm{d}t}_{\text{Diffusion contribution}}
\label{eq:inst_Lift}
\end{aligned}
\end{equation}
In the above equation, the first integral from left is the first $x$-moment of the vorticity field ($A$), with $\omega_z(t)$ as the instantaneous spanwise vorticity field.
The second integral from left is the contribution from the viscous term (diffusion), where $\nu$ is the kinematic viscosity and $\rho$ is the fluid density.

Applying Taylor's hypothesis (as introduced above for the wake reconstruction), $\mathrm{d} x = U_\infty \mathrm{d}t$, one can transform the spatial derivative in the left-side integral of \Cref{eq:inst_Lift} into a temporal one. 
Moreover, by interchanging the left-side integral in \Cref{eq:inst_Lift} with the time derivative (using Leibniz integral rule), one can re-write \Cref{eq:inst_Lift} as follows:
\begin{equation}
L_{circ}(t) = \rho U_\infty \int_0^t \left[ \int_0^h u\omega_z(t)\mathrm{d}y +  \int_0^h\nu \left( \frac{\partial^2u(t)}{\partial x^2} + \frac{\partial^2u(t)}{\partial y^2} \right)\mathrm{d}y \right] \mathrm{d}t
\label{eq:inst_Lift2}
\end{equation}
Therefore, the change in the lift in time $\delta\tau$ can be expressed accordingly:
\begin{equation}
\delta L_{circ} = \rho U_\infty  \delta\Gamma
\label{eq:delL}
\end{equation}
where the corresponding change in the circulation $\delta \Gamma$ is given by:
\begin{equation}
\delta\Gamma =  \int_0^h u\omega_z(t)\mathrm{d}y +  \int_0^h\nu \left( \frac{\partial^2u(t)}{\partial x^2} + \frac{\partial^2u(t)}{\partial y^2} \right)\mathrm{d}y
\label{eq:delGamma}
\end{equation}

Since at the beginning of the unsteady motion the lift is unknown, we shall refer to the estimated lift component as an increment in the circulatory lift \cite{Stalnov2015,BenGida2016} that is generated from the beginning of the motion. 
Based on \Cref{eq:inst_Lift2}, the cumulative circulatory lift at time $t$, $\Delta L_{circ}(t)$, is computed accordingly:
\begin{equation}
\Delta L_{circ} (t)= \rho U_\infty \int_0^t \zeta(t)\mathrm{d}t = \rho U_\infty \Gamma(t)
\label{eq:dL-circ}
\end{equation}
with the vorticity flux term $\zeta(t)$ being expressed as follows:
\begin{equation}
\zeta(t) = \int_0^h u_c\omega_z(t)\mathrm{d}y +  \int_0^h\nu \left( \frac{\partial^2u(t)}{\partial x^2} + \frac{\partial^2u(t)}{\partial y^2} \right)\mathrm{d}y
\label{eq:zeta_vs_time}
\end{equation}
Here, $u_c$ is the advection velocity at which the characteristics of the wake collectively travel downstream.
The cumulative circulatory lift coefficient at time $t$ is therefore expressed as:
\begin{equation}
\Delta C_{l_{circ}}(t) = \frac{2}{cU_\infty} \int_0^t \zeta(t)\mathrm{d}t = \frac{2\Gamma(t)}{cU_\infty}
\label{eq:dCl-circ}
\end{equation}

As detailed in \Cref{getWAKEfunctions}, the getWAKE Toolbox suggests two options for computing the cumulative circulatory lift coefficient in the wake. 
The first option is based on Panda and Zaman method \cite{Panda1994}, where the vorticity flux $\zeta(t)$ defined by \Cref{eq:zeta_vs_time} is computed for each individual PIV velocity map obtained in the wake behind the body (as function of time), after applying a threshold on the vorticity contours.
For each velocity map (or time $t$), the advection velocity is defined as the mean streamwise velocity component along the $x$-direction, $u_c\approx\langle{u}\rangle_x(y)$, and the spanwise vorticity field is estimated as the local mean spanwise vorticity along the $x$-direction, $\omega_z(x,y,t)\approx\langle{\omega_z}\rangle_x(y,t)$. 
Moreover, for each velocity map, the second order derivatives of $u$ are computed using a least squares differentiation scheme and then approximated with their local mean value along the $x$-direction; $\partial^2u(x,y,t)/\partial x^2\approx\langle{\partial^2u/\partial x^2}\rangle_x(y,t)$ and $\partial^2u(x,y,t)/\partial y^2\approx\langle{\partial^2u/\partial y^2}\rangle_x(y,t)$.
All the above vectors, representing each instantaneous velocity map, are then integrated over the $y$-direction to yield the vorticity flux $\zeta(t)$, as presented in \Cref{eq:zeta_vs_time}, and the time variation of the cumulative circulatory lift coefficient in the wake (see \Cref{eq:dCl-circ}).
Using the option described above, the computation of the cumulative circulatory lift coefficient for a sequence of velocity maps does not require \textit{a priori} information of the reconstructed wake.
Nevertheless, if needed, in the getWAKE Toolbox, the user can choose to compute the vorticity flux $\zeta(t)$ directly from the reconstructed wake signature and not from individual velocity maps; with or without a threshold applied on the vorticity contours. 
In doing so, the advection velocity is approximated as the freestream velocity, $u_c\approx U_\infty$ and the spanwise vorticity field array is taken directly from the reconstructed wake, $\omega_z(x,y)$, where the $x$-direction and time $t$ are interchangeable by applying $\mathrm{d}t=U_\infty\mathrm{d}x$.
The second order derivatives of $u$ are computed using a least squares differentiation scheme directly from the reconstructed wake; $\partial^2u(x,y)/\partial x^2$ and $\partial^2u(x,y)/\partial y^2$.
All the above arrays, representing the full reconstructed wake map, are then integrated over the $y$-direction (at each $x$ location) to yield the vorticity flux $\zeta(t)$, as presented in \Cref{eq:zeta_vs_time}.

In the second option, the cumulative circulatory lift coefficient in \Cref{eq:dCl-circ} is computed based on a direct summation of the circulation values throughout the reconstructed wake signature. 
Thus, $\zeta(t)$ is computed as the integral over the spanwise vorticity field of each individual PIV velocity map, accordingly:
\begin{equation}
\zeta(t) = \int_0^l \int_0^h \omega_z(x,y,t)\mathrm{d}x\mathrm{d}y
\label{eq:zeta_option2}
\end{equation}


\newpage
\bibliographystyle{elsarticle-num} 
\bibliography{bibliography}






\end{document}